\documentclass[12pt]{article}
\usepackage{graphicx,psfrag,epsf}
\usepackage{enumerate}
\usepackage{url} 

\usepackage{amsmath}
\usepackage{amsfonts}
\usepackage{amsthm}
\usepackage{amssymb}
\usepackage{multimedia}
\usepackage{verbatim}
\usepackage{times}
\usepackage[T1]{fontenc}
\usepackage{booktabs}
\usepackage{tabularx}
\usepackage{colortbl}
\usepackage{graphicx}
\usepackage{amsmath}
\usepackage{amsfonts}
\usepackage{amsthm}
\usepackage{fancyhdr}
\usepackage{framed}
\usepackage{natbib}
\usepackage{ulem}  
\usepackage{algorithmic}
\usepackage{algorithm}
\usepackage{subcaption}
\usepackage{caption}

\newcommand{\blind}{0}

\begin{document}

\def\spacingset#1{\renewcommand{\baselinestretch}%
{#1}\small\normalsize} \spacingset{1}


\if0\blind
{
  \title{\bf Teaching Split Plot Experiments With a Boomerang Tin.}
  \author{Thomas Muehlenstaedt \thanks{
    The authors gratefully acknowledge the support of W. L. Gore \& Associates}, Maria Lanzerath \hspace{.2cm}\\
    W. L. Gore \& Associates \\
}
  \maketitle
} \fi

\if1\blind
{
  \bigskip
  \bigskip
  \bigskip
  \begin{center}
    {\LARGE\bf Boomerang tin}
\end{center}
  \medskip
} \fi

\bigskip
\begin{abstract}
	This article presents an example on how to teach split-plot experimental designs based on a hands-on exercise. This is a toy called boomerang tin which utilizes a rubber band to store and release energy. The set up and mechanisms of the hands-on example are explained followed by a description of how it can be leveraged in teaching split plot DoE. An actual design is set up, analyzed and discussed to provide reference for its usage in class.

\end{abstract}

\vspace{0.5cm}

Keyworks: teaching aids, design of experiments, restricted randomization, hands on teaching examples, analysis of variance
\section{Introduction}

Design of experiments (DoE) are a well known tool in industry. Training classes in DoE are widely available in university and corporate settings and often combine a mixture of lecture, software demonstrations and hands-on exercises. Hands--on exercises are particularly valuable for students in training classes because they give them a chance to execute a real experiment and gain a deeper understanding of fundamental DoE concepts, and how they apply to real experiments. We have successfully used several different types of hands-on exercises in our work as DoE class trainers. 

Some DoE concepts related to split-plot experiments are more difficult to teach in existing hands-on exercises. In this article, a new hands--on exercise for teaching split-plot designs is presented. It uses a toy called a boomerang tin. The exercise makes use of the boomerang tin's property to first roll forward and then backwards again. This is caused by a rubber band with a weight attached to it, as the rubber band stores energy. The boomerang is easy and cheap to build from items that can be purchased at a building supplies store.

There are other hands--on exercises for teaching statistical DoE that have been presented in the literature. One of the earliest descriptions of a hands--on experiment is Fisher's Tea Tasting Experiment (\cite{Fisher1960DoE}). Use of a small catapult can introduce  students to DoE methods. \cite{Luner1994Catapult} uses the catapult to demonstrate response surface design and process optimization while in \cite{Antony2002Catapult} catapults are used to perform screening experiments and to explore interactions. Another popular exercise are paper helicopters (\cite{Box1992Helicopter}, \cite{Annis2005HelicopterAmstat}, \cite{Box99StatsAsCatalyst}), used to explain fractional factorial designs and response surface methods. 
\cite{Montgomery2009DoE} describes a cake baking experiment that is practical and easy to understand. None of these examples connects to the teaching of hands--on split--plot experiments. This fact motivated our work.

While these experiments have been used for teaching statistics and some types of DoE, we introduce a practical example that mimics the typical application case for a split-lot experiment, and show how it can be used in a hands-on exercise in a DoE training class. 

Split--plot experiments are a special case of blocked experiments and have their origin in agricultural applications. \cite{Kowalski2007TutorialSplitPlot} and \cite{Jones2009SplitPlotDesigns} give good introductions into split--plot designs. They have become more and more common in industry, as they allow to handle different randomization schemes for different types of factors. It is more efficient to use different randomization schemes if the levels of one or more factors are not as easy to change as for other factors. The boomerang tin offers both, easy and hard to change factors and hence an appropriate exercise that can clearly demonstrate key principles and support the effective teaching of split--plot designs.
	
This article is organized as follows: We start with a brief introduction to split--plot experimental design and analysis, followed by a description of the boomerang tin. The next section presents a split--plot screening experiment, suited for a training class, with 4 factors from the boomerang tin, followed by a conclusion. 
	
\section{Split--Plot Experiments}

Split--plot designs are a special category of designed experiments in which blocking of one or more factors is needed. In industrial experiments, randomization is a fundamental principle for running experiments. Randomization reduces the risk of making wrong conclusions due to unknown or uncontrollable factors. Fully randomized designs require that factor settings can be easily changed with each run as prescribed by the design, like a speed of a machine or an air pressure.
	However, sometimes factor levels are not easy to change. Imagine a temperature with two different settings of 100 and 150 degrees Celsius on an oven where heating up and cooling down between temperature changes is time consuming, it will be hard or even impractical to change the temperature between every single experimental run in the experiment. Other examples of hard-to-change factors would be a tooling part requiring de-assembling and assembling machines, or chemical mixtures, where a chemical lot needs to be of a given size because of minimum mixing amounts, but only a small portion of the mixture is actually used in the experiment. In these cases, a fully randomized DoE quickly becomes a challenge due to practical constraints, or cost. Therefore randomization will be restricted for the hard--to--change factors while full randomization is still feasible for the easy--to--change factors in the experiment at the same time.\\
	Having two different randomization schemes can be understood as having two different designs nested in each other. The "outer" design contains the factors which are too hard to be fully randomized. The "inner" design deals with the easy-to-change factors and from that perspective, the outer design runs are considered as random blocks.
	Split Plot DoEs can be set up using optimal design theory, e.g. in JMP's Custom Designer platform (\cite{JMP11}) or Minitab (\cite{Minitab18}).
	
	Suppose that we have a response variable $(Y)$ measured on each of $n$ runs in the experiment, with $Y_{ij}$ corresponding to a specific realization. We use $i=1$ to $r$ to refer the number of random blocks which are sometimes called whole plots. We use $j=1$ to $n_i$ to represent the number of runs for each whole plot so we have $n=n_i r$. The runs are split equally among random blocks. In order to represent the nested structure of the split--plot design, a second error term for the outer design is introduced into the statistical model (\cite{Goos2002optBlockedSplitPlotDoEBook}). Let $\beta$ be the coefficients to estimate, and $\epsilon$ and $\gamma$ be error terms:
	\begin{equation}
		Y_{ij} = f(w_i, s_{ij}) \beta + \gamma_i + \epsilon_{ij}.
	\end{equation}
	Here $w_i$ represents the hard to change factors (so called whole plot factors) and $s_{ij}$ represents the easy to change factors (so called subplot factors). Splitting up the function $f$ into one part $f_w$ only depending on the whole plot factors and another part $f_s$ for the subplot factors and the interactions between sub- and whole plot factors results in the following equivalent description:
	
	\begin{eqnarray*}
			Y_{ij}  & =  & f_w(w_i) \beta_w + \gamma_i \\
			& + & f_s(w_i, s_{ij}) \beta_s +  \epsilon_{ij}.
	\end{eqnarray*}
	Writing the model equation in this way reveals why a split--plot experiment can be understood as doing two nested experiments.	
	The two different randomization schemes are represented by those two error terms. Both are assumed to be normally distributed, centered, and uncorrelated to each other with variances $\sigma_{\gamma}$ and $\sigma_{\epsilon}$.
	Index $i$ refers to the different whole plots, index $j$ refers to the subplot factors.
	
	Although the two error terms are assumed to be uncorrelated, there is a non--diagonal covariance structure resulting from the model. Hence the covariance of one whole plot is given by:
	\begin{equation*}
		\Gamma = \begin{bmatrix}
		\sigma_{\epsilon}^2 + \sigma_{\gamma}^2 & \sigma^2_{\gamma} & \dots & \sigma^2_{\gamma} \\
		\sigma^2_{\gamma} & \sigma_{\epsilon}^2 + \sigma_{\gamma}^2 & \dots & \sigma_{\gamma}^2\\
		\vdots & \vdots & \ddots & \vdots \\
		\sigma^2_{\gamma} & \sigma^2_{\gamma} &\dots & \sigma_{\epsilon}^2 + \sigma_{\gamma}^2\\ 
		\end{bmatrix}
	\end{equation*}
	 Then the covariance matrix of the overall design is a block diagonal matrix:
	\begin{equation*}
		V = \begin{bmatrix}
		\Gamma & 0 & \dots & 0\\
		0 & \Gamma & \dots & 0\\
		\vdots & \vdots & \ddots & \vdots \\
		0 & 0 & \dots & \Gamma
		\end{bmatrix}.
	\end{equation*}
This small restriction in randomization actually leads to significant consequences for the generation and analysis of split--plot designs:
\begin{itemize}
  \item In the design generation, the corresponding optimality criteria for optimal designs get more complex.	
  \item In the analysis, ordinary least squares regression is no longer valid in general. There are Split Plot designs, where the analysis does not need to be changed to obtain correct p-values (\cite{JonesGoos2011EquivalenEstimationSecondOrderSplitPlotDoE}). However this is not true for all numerically obtained Split-Plot designs, and generalized least squares have to be used instead. As the covariance is only known up to the variances $\sigma_{\epsilon}^2$ and $\sigma_{\gamma}$, these have to be estimated by means of variance component analysis, e.g. restricted maximum likelihood estimation, hence software is needed to obtain the model parameter estimates. (\cite{GoosJones2011CaseStudyBook}, 
	\cite{Goos2002optBlockedSplitPlotDoEBook}).
\end{itemize}
	
	One consideration when setting up any designed experiments is to check: are there enough degrees of freedom available to estimate all terms of the model
	and still have enough degrees of freedom for a reliable error estimate?
	The split-plot structure comprises a change here: in a fully randomized design, the 
	error 
	degrees of freedom are the same for all terms, whereas in a split-plot structure there are quite different numbers of 
	error 
	degrees of freedom available to estimate the model terms for the whole plot factors and for estimating terms on the subplot level. This leads to the necessity to carefully select both the number of whole plots and the size of the whole plots (for simplicity, we assume uniform size of the whole plots). This will be studied in more detail in section 3.3.

\section{Using the boomerang tin in teaching a DoE class}
\subsection{Split Plot overview}
Before going into DoE details for the boomerang tin, we would like to shortly descibe our approach to setting up a Split Plot DoE. This is not intended as a complete treatment of Split--Plot DoE theory, for a more rigourus treatment please refer to \cite{Goos2002optBlockedSplitPlotDoEBook}. 
To set up a Split--Plot DoE in practice, the following steps can be taken. These steps assume a given model to be estimated as well as a definition of which factors are hard to change (whole plot factors) and which are easy to change (subplot factors). Furthermore, the concept of counting degrees of freedoms is assumed to be known.
\begin{enumerate}
	\item Set up the model and count degrees of freedom (df) needed for the whole plot experiment 
	\item Set up the model and count degrees of freedom needed for the sub plot experiment
	\item Choose suitable numbers for the number of runs (total sample size) and the number of whole plot runs. 
	\item Set-up a design with a software tool that offers numerically optimized split--plot experiments.
	\item Evaluate the design, and if not good enough yet, go back to step one again and re--iterate the process for improvement.
\end{enumerate}


We found that for teaching the design planning, it is key to understand the concept of counting degrees of freedom. So we encourage the students to create a table of the degrees of freedom as part of the exercise (see Table \ref{table:CountDF} for an example for the boomerang tin).  As the split--plot structure can be understood as nesting two  experiments, counting degrees of freedom also happens in two steps. 

As our rule of thumb, at least 4 degrees of freedom for the whole plot error estimate are aimed at and at least 5 (and idealy no more than 10) degrees of freedom for the subplot error estimate.

\subsection{Description of a boomerang tin and its factors}

\begin{figure}
\centering
\begin{subfigure}{.49\textwidth}
	\centering
	\includegraphics[width=.92\linewidth]{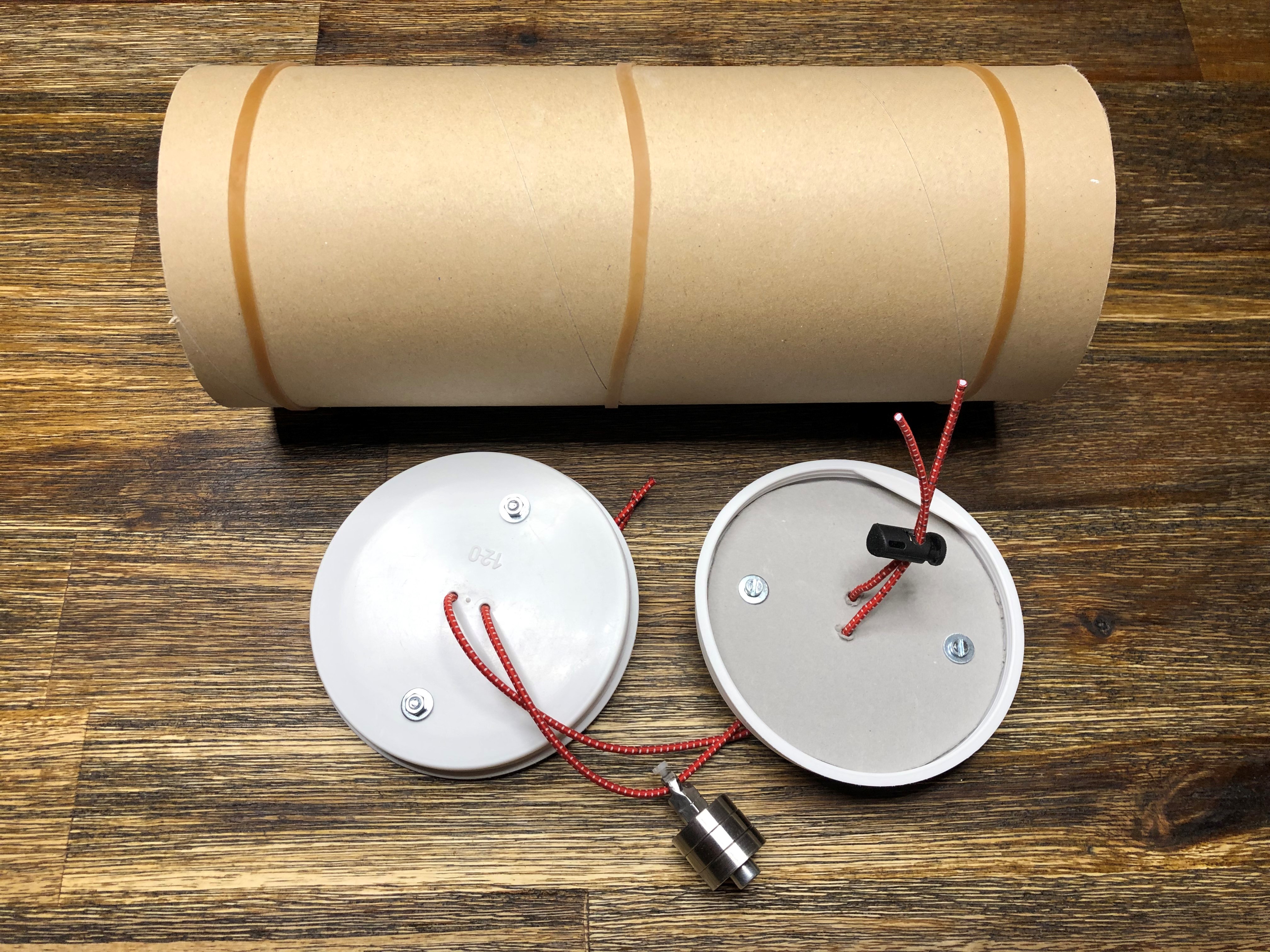}
\end{subfigure}
\begin{subfigure}{.49\textwidth}
	\centering
	\includegraphics[width=.94\linewidth]{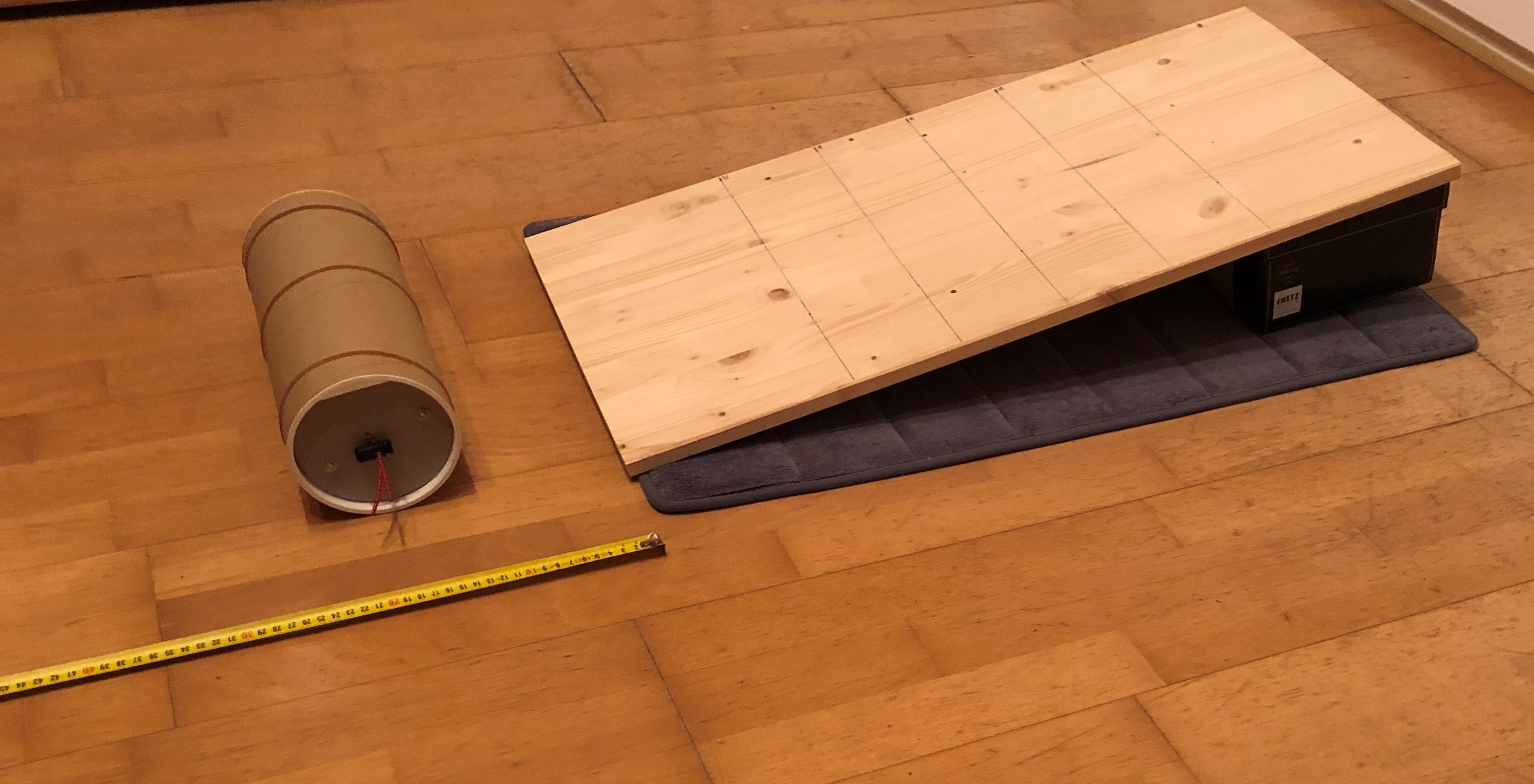}
\end{subfigure}

\caption{Boomerang tin, disassembled (left) and assembled (right)}
\label{fig:BoommerangTinDescription}
\end{figure}

The boomerang tin is a toy described in  \cite{Press2002LittleGiantScience}, p. 160, and originally made to explain gravity. 
See Figure \ref{fig:BoommerangTinDescription} for a picture. In order to construct a boomerang tin, a cylinder along with some lids are needed. Rubber bands can be taken from a textile supply as well as so called cord stoppers and metal screw nuts of different size as weight. 
It is important to use high quality rubber bands, as for inferior quality, the rubber band will degrade already within one day of experimentation. Therefore we recommend to use rubber bands as used in textile industry. To our experience those work fine.
Also, a ruler can help control the tension of the rubber band by measuring its length outside of the tin. All these items can be found in a well equipped building center. 

When rolling the boomerang tin, the weight (e.g. a simple screw might work) inside prevents the rubber band from moving together with the tin. The rubber band does not turn at all but builds up energy instead. Once the boomerang tin stops rolling forward, the energy stored in the rubber band  gets resolved by causing the boomerang tin to move backwards.The key mechanism that makes the tin work is the ability of the rubber band to store energy.\\

For the planning of a designed experiment, we first select the factors to vary. There are two groups of factors in the boomerang tin example: one group around the tin setup, and another group around the rolling of the tin. For the tin setup, we choose different weights of the screw (factor $x_1$, assumed to be categorical). Another factor is the tension of the rubber band (factor $x_2$, assumed to be continuous). The rubber band holder is placed outside the white plastic lids. The tension can be set by changing the length of the rubber band outside the tin. 
The third potential factor is whether the rubber band is pre--twisted or not (factor $x_3$, two level categorical). 
	
The nut weight ($x_1$) is chosen to be categorical in this example, only two kinds of weights were built for the experiment. In principle it could also be treated as a continuous factor. The weight could be varied e.g. by utilizing screws of different diameter in a simple setting. The screw weight is a hard--to--change factor, as the complete boomerang tin needs to be disassembled for each change. This quickly becomes tedious, and, even more important, can introduce a lot of variability into the system and its responses. If not done carefully, the whole--plot variance will increase. This suggests the nut weight to be a natural hard--to--change factor.

For factors around the rolling of the boomerang tin: the tin starts travelling down a ramp of a defined slope. In fact this slope factor is set through the height of the ramp at start (factor $x_4$, assumed to be continuous). See Figure 2 for the experimental set up of the ramp.

\begin{figure}
	\label{fig:ramp}
	\centering
	\includegraphics[width=0.75\textwidth]{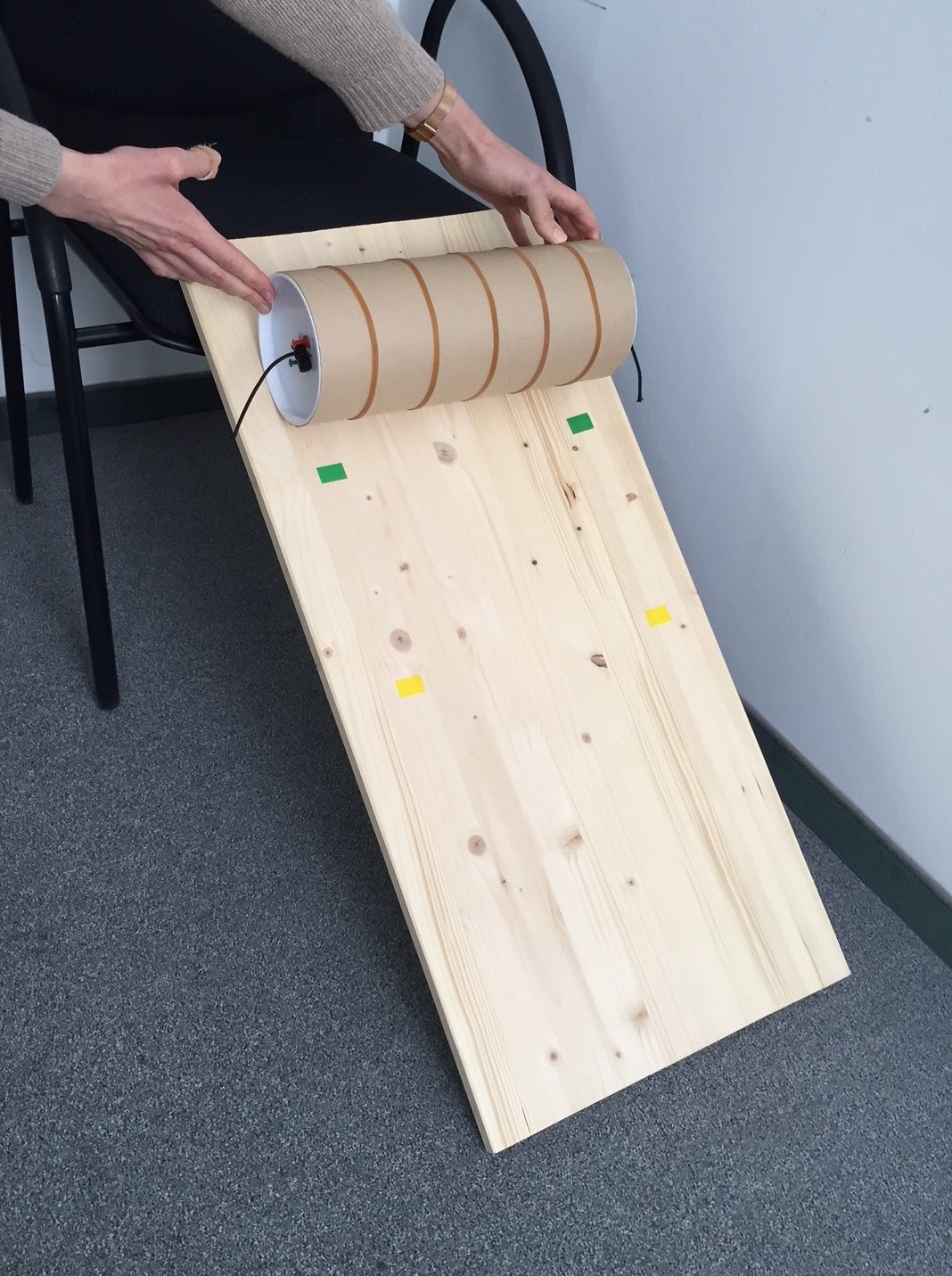}
	\caption{Experimental setup of the starting ramp rolling the boomerang tin.}
\end{figure}

There are two outputs considered: the distance the boomerang tin rolls forward ($y_1$) and the distance it rolls backwards ($y_2$).
Both are measured by a 5 meter long ruler on the floor. For easier reading, some small adhesive stripes as visual guides can be added in the range of interest. Please see Figure \ref{fig:BoommerangTinDescription} for an image of the measurement process. In order to get reliable readings, it is recommended to split tasks in a students' team: while one releases the boomerang tin, another team member reads both outputs while the boomerang tin is moving. 
 These two outputs tend to depend on the factor settings in an opposite manner, and both should be maximized at the same time. We believe that the toy provides more learning opportunities when the boomerang effect is large, and also gives an interesting challenge when analysing the data.


There are several noise factors associated with this experiment. For example, consider the floor surface (tile or carpet). That can have a high impact on the rolling behavior. In case of a carpet, also the direction of its fibers affects the rolling properties. Another unwanted behaviour, or noise, occurs if the rubber band is not fixed strong enough at either end: Eventually it starts turning in an uncontrolled manner. This can be avoided by adding double sided duct tape, so that the rubber band holder cannot move as easily, or by fixing the rubber band outside the tin. We summarize the factors that we chose for the exercise in Table 1, which lists the factors, their types, the possible levels and whether or not they are hard--to--change factors.

One could add further input factors, e.g. the length of the starting ramp as another easy--to--change factor, or its surface. Different roll diameters could serve as another hard--to--change factor, or different elasticities of the rubber band. In order to keep the experiment easy to execute as a hands--on exercise in class, we typically keep the number of factors small. We found the listing of factors in Table 1 is appropriate to stay within a reasonable time frame in a classroom setting, and at the same time allows the students to learn the key concepts of split--plot designs. Using more than 4 factors will only increase the complexity with no value added.

\begin{table}
\begin{center}
\begin{tabular}{|r|c|c|c|c|c|}
\hline
Factor & Description        & Lower level & Upper level & Type & Hard to change?\\
\hline
$x_1$ & nut weight               & low  & high & categorical & yes\\
$x_2$ & tension of rubber band   & low  & high & continuous  & no\\
$x_3$ & rubber band twisted      & no    & yes   & categorical & no\\
\hline  
\end{tabular}
\caption{Experimental factors and their descriptions.}
\end{center}
\end{table}


\subsection{DoE set up}

Table \ref{table:WP_model} below shows the counting results for degrees of freedom in the first step, the whole--plot experiment:

\begin{table}[!ht] \begin{center}
\begin{tabular}{l | r| r}
Whole Plot Effect(s) &  Levels & DF\\
\hline\hline
screw weight: $x_1$ & 2 & $ 2-1=1$\\
\hline
overall mean		& & 1\\
\hline
error estimate		& &$\geq 4$\\
\hline\hline
Sum		&	& $\geq 6$\\
\end{tabular} \end{center}
\caption{Set up of the whole--plot model for counting degrees of freedom}
\label{table:WP_model}
\end{table}	

 2 df are used for estimating the whole--plot effects. The authors recommend to allow minimum 4 df for the error estimation wherever possible for avoiding even lower power values. As a consequence, 6 or more whole plots are recommended for this example.

The second step is to count the DF for subplot level, as well as for the total experimental design size.
\begin{table}[!ht] \begin{center}
\begin{tabular}{l | r| r}
	
Sub Plot Effects          &Levels & DF\\
\hline
takeover from Whole Plot    &     & 6 \\
Sub Plot main effects:      &     &   \\
$x_2$ (tension rubber band) & 2   & 1 \\
$x_3$ (rubber band twisted) & 2   & 1 \\
$x_4$ (position ramp)       & 2   & 1 \\
2--factor interaction within Sub Plot:      &     &   \\
 $x_2 x_3$                  & 4   & 1 \\
 $x_2 x_4$                  & 4   & 1 \\
 $x_3 x_4$                  & 4   & 1 \\
2--factor interaction Whole Plot to Sub Plot: &     &   \\
 $x_1 x_2$                  & 4   & 1 \\
 $x_1 x_3$                  & 4   & 1 \\
 $x_1 x_4$                  & 4   & 1 \\
 \hline 
 sum (model DFs)            &     & 15 \\
 error estimation ($\geq 5$ DFs)& & proposal: 9 \\
\hline\hline
Sum (total number of EUs)	&     & 24\\
\end{tabular} \end{center}
\caption{Counting degrees of freedom for each effect in the model.}
\label{table:CountDF}
\end{table}

Note that the degrees of freedom for any interaction between a whole--plot factor and a sub--plot factor are counted on the sub--plot level.
As the easy--to--change factors are usually not that "expensive", a minimum 5 df for the error estimation will often be used, which ensures an acceptable level of power for the parameter estimates. 

Knowing that 6 whole plots and in total 20 or more runs are the minimum needed, a total run size can now be determined. Some different designs can be tried out by varying the number of sub--plot error degrees of freedom, as well as the whole--plot experiment size. In the first try, our best guess, we use $n=6 \cdot 4= 24$ runs, such that each whole--plot would have equal size of 4. This number of 4 allows the chance, for each whole--plot block, to be a half--fractional factorial for the sub--plot factor experiment. Taking such kind of reasonings into consideration, in tendency leads to more balanced designs. Other possible choices might be 8 whole plots and 32 runs (leading to powers close to 1 for the sub--plot factors and a quite large overall design) or maybe 8 whole plots and 24 runs, leading to whole plots of size 3, each.

To create the design, we used JMP's Custom Design platform, which uses a design generation algorithm. Depending on software, there are various options to be set such as optimality criterion (here D--optimality), number of starts of the search, or the ratio of the two error variances to be estimated (see the JMP 11 Design of Experiments Guide, \cite{JMP11DoEGuide}, or the most recent version, for explanations).

After generating the design, its properties will be analysed, e.g. the power for the effect estimates: for a signal to noise ratio of 1 the power should be minimum $80\%$, as a rule of thumb. This is usually not achieved for the whole plot effects. A good way to achieve higher power is to use more whole plots. This is often impossible from a practical point of view, and eventually would diminish the reasons for running a split--plot design which is to reduce the frequency of changes for the hard--to--change factors. Hence the lower power for the whole plot factor is tolerated.
Other meaningful analysis checks regarding design evaluation are correlation matrices, variance inflation factors (or similar methods), or prediction variances. Given those are on an acceptable level, the design can be used.

\subsection{DoE Analysis}
\begin{figure}
	\label{fig:DoE}
	\centering
	\includegraphics[width=0.75\textwidth]{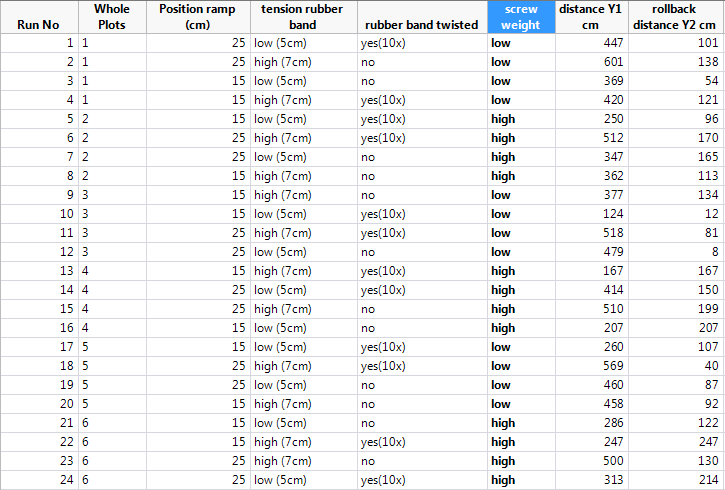}
	\caption{DoE table including results.}
	\label{Fig:Dataset}
\end{figure}

We now demonstrate the analysis with a real sample dataset
which is provided in Figure \ref{Fig:Dataset}. 
A split--plot design with n=24 experimental units was executed, and real response values for both forward and backward distance were measured and collected. This section describes different aspects of the analysis results and conclusions. We will share the results of the analysis using JMP software. The analysis could be done in any software that performs generalized least squares regression and variance component analysis.

The error structure of a Split Plot DoE is the one of a random blocks design. To analyse the design properly, the table needs to be complemented with a column containing the whole--plot blocks. This column will be added as a random block effect to the model to fit.
Due to the two error terms, the generalised least squares estimator has to be applied, and also a method for estimating the variance components has to be chosen (here REML methods are applied). 

The analysis results are given in a sequence of figures.
Figure 3 shows the overall fit of the linear model for both responses. Overall, for the forward distance $Y_1$, an $R^2$ of $0.89$ was achieved together with a root mean square error of 60.3 cm. For the rollback distance, the $R^2$ is 0.52 with a root mean square error of 53.9 cm. The lower $R^2$ can been recognized in the plot for $Y2$, as the observed points (in black) scatter more around the red predicted line than in the plot for $Y1$. These graphs also help check the assumptions for the linear model that was used. No objections can be seen.

For both models, the significance test of the overall model turns out to be highly significant, indicating that the models in fact do explain output variation. 

\begin{figure}[ht]
\label{fig:ActualByPredicted}
  \centering
  \includegraphics[width=\textwidth]{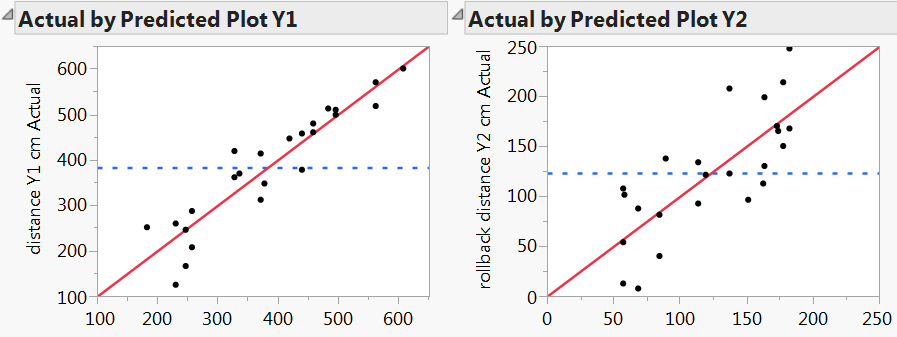}
\caption{Actual by predicted plots for both outputs $Y1$ and $Y_2$, forward and backward distance.}
\end{figure}

Testing the input factors for statistical significance leads to Figure 5. For the forward distance $Y_1$, all main effects are significant but no interaction, whereas for the rollback distance, $Y_2$, only the screw weight is significant. It would be reasonable to expect that for $Y_2$ at least the tension of the rubber band could also have an impact, but the related p-value is clearly insignificant, which might indicate that levels have  been chosen too narrow.

\begin{figure}[ht]
\label{fig:Fig2345}
  \centering
  \includegraphics[width=\textwidth]{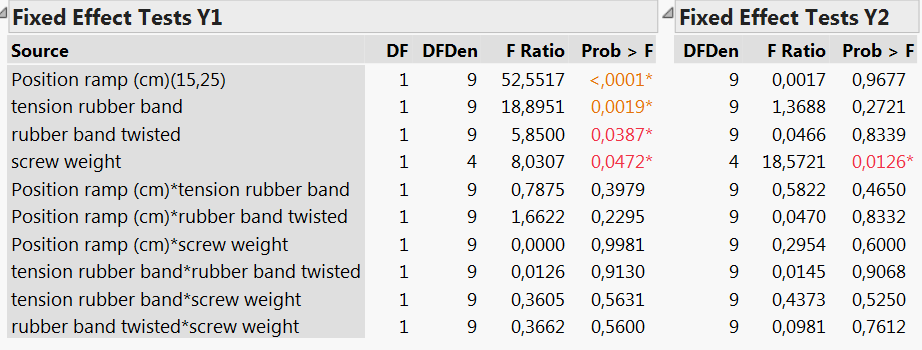}
\caption{Fixed effect tests for both outputs $Y_1$ and $Y_2$.}
\end{figure}
The prediction profiler results in Figure 5 can help choosing a factor combination setting that optimizes the predicted responses simultaneously. In our case it is favorable to use a heavy screw weight in order to assure a long rollback distance and to achieve a high distance for $Y_1$ as well.
\begin{figure}[ht]
\label{fig:PredictionProfiler}
  \centering
  \includegraphics[width=\textwidth]{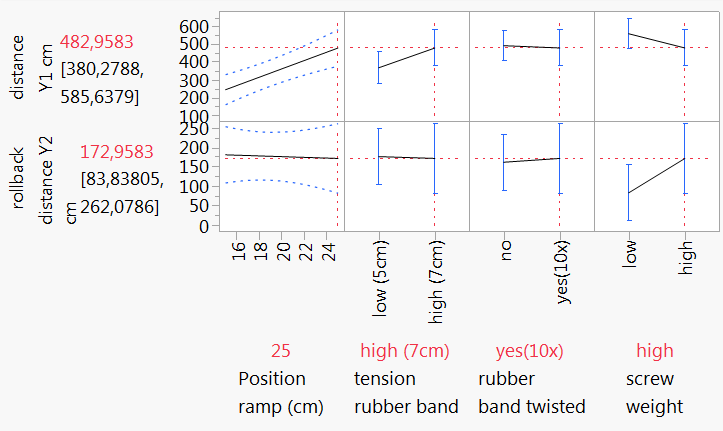}
\caption{Prediction profiler for both outputs $Y1$ and $Y_2$.}
\end{figure}

We check the model assumptions (linearity, independence, normality, homoscedasticity) of the fitted linear model in order to decide if the model analysis is valid or if some other model is needed. This step should never be forgotten. A residual by predicted plot is used to check the latter two. If e.g. the model assumption of homoscedasticity is fulfilled, the residuals should scatter around the zero line randomly with no funnel pattern visible as can be seen in Figure 6. In general, the residuals have to randomly scatter around zero.

\begin{figure}[ht]
\label{fig:ResidualByPredicted}
  \centering
  \includegraphics[width=\textwidth]{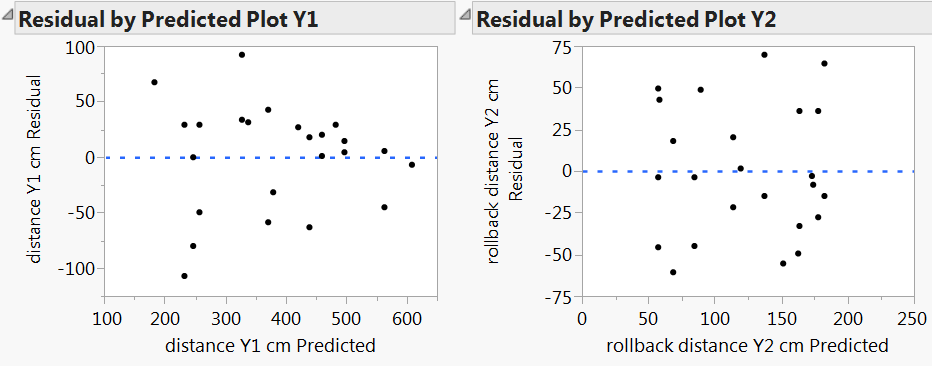}
\caption{Residual plots for both outputs $Y_1$ and $Y_2$.}
\end{figure}

\subsection{Some practical hints for using the exercise in DoE training class}
When using a hands--on example in teaching, concise instructions for the exercise and a clear experimental objective should be formulated. The overall objective of the experiment could be one of the following statements:

1) Evaluate which of the main effects and two-factor interactions of the varied factors have an impact on the two outputs, respectively. Or:

2) Find a best setting to maximize both responses at the same time.\\

The range finding and factor screening process quickly gets time consuming, so the time to execute the exercise can be reduced drastically if the class instructor predetermines some initial factor ranges for the students to use. Some factor settings will lead to no boomerang effect. As a result, factor ranges should be selected in a way that ensures the boomerang tin will function as intended so that the students do not waste time in finding factor ranges for the experimental setup. This allows them to spend more time on learning the split--plot design concepts. Based on that knowledge, the factor levels to be used in the exercise could be given to the students beforehand.

In class, the students are split into groups of 3-5 people so that every participant gets a chance take an active part in the exercise. First, the students are asked to make themselves familiar with the set up and to run some initial  trials. To keep the time short, the number of pre--trials could be restricted to five or fewer runs. This will also make them familiar with the measurement process for the two responses, the distances. This does require some consideration. The instructor could give the students some time to determine how they will do the measurements to reduce the impact of measurement variability and retrieve reliable results right from the beginning.
\bigskip

We found it works well to give the students a set of steps to follow in the exercise and suggest a plan like the list below. The instructor can use the timing to help keep the small group teams on track and adjust the timing as needed.
\begin{enumerate}
	\item Determine goals of experiment (5 minutes)
	\item Do 5 or fewer test runs to become familiar with the boomerang tin (10 minutes)
	\item Plan the experiment and create a run sheet (10 minutes)
	\item Execute the experiment (15 minutes)
	\item Analyze the data and discuss conclusions (15 minutes)
\end{enumerate}

As a potential summary, the trainer could show a visual analysis of the combined designs, and show how the results of the different teams compare to each other.

\section{Conclusion}

The shown example of a hands-on training exercise is especially suitable for teaching the concept and application of split--plot experiments in DOE classes. Split--plot designs have factors which vary in their difficulty to change their levels. While catapults and paper helicopters offer nice exercises for teaching DoE concepts, they do not lend themselves very well toward split--plot experiments. We believe the boomerang tin is a suitable alternative due to its potential to add hard--to--change factors. 

As a hands--on exercise quickly takes a significant length of time in class, a thorough preparation by the teacher dramatically helps manage time while providing effective learnings. Hardly anything is more frustrating for all parties than an exercise that does not work. We recommend to fix factors in advance and to give precise tasks and instructions. Separating the exercise into smaller chunks, potentially with a fixed timing, helps keeping the class moving forward.
Overall, based on the authors experience, this exercise can be covered by students in about 1 to 1.5 hours.

In addition, the boomerang tin has results in multiple responses so it is a good example for teaching simultaneous optimization of multiple responses.

In case there are practical questions around handling the authors would be happy to support and give further advise.

  \bibliographystyle{chicago}
\bibliography{LitMuehlenstaedt}
\end{document}